\documentclass[letterpaper]{jpconf}
\usepackage{graphicx}

\def\bea{\begin{eqnarray}}
\def\eea{\end{eqnarray}}

\begin{document}
\title{Correlations from p-p collisions at $\sqrt{s} = 200$~GeV}

\author{R Jefferson Porter and Thomas A Trainor (STAR Collaboration)}

\address{CENPA 354290 University of Washington, Seattle WA 98195 USA}

\ead{trainor@hausdorf.npl.washington.edu}

\begin{abstract}
We describe a comprehensive survey of two-particle correlations from p-p collisions at $\sqrt{s} = 200$ GeV. Correlations on transverse rapidity $y_t$ are separated into a soft component (string fragments) and a hard component (parton fragments). Fragment correlations on transverse rapidity $y_t$ are compared to fragmentation functions on logarithmic momentum variable $\xi$. Cuts on transverse rapidity isolate angular correlations on $(\eta,\phi)$ for the two components. Hard-component angular correlations are compared to results from conventional leading-particle jet analysis. Angular correlations for low-$p_t$ fragments, corresponding to low-$Q^2$ parton scattering, reveal a large asymmetry about the jet thrust axis.
\end{abstract}.

\section{Introduction}

QCD theory predicts that an abundance of low-$Q^2$ scattered gluons (minijets) should be produced in relativistic nuclear collisions at RHIC energies. Those gluons are believed to drive formation of the colored medium in heavy ion collisions~\cite{theor0,theor1}. If so, we may discover remnants of low-$Q^2$ ($Q/2 \sim$ 1 - 5 GeV) partons in the correlation structure of final-state hadrons. Initial studies of two-particle correlations in p-p collisions emphasized momentum subspace $(\eta,\phi)$ (pseudorapidity and azimuth)~\cite{etaphicorr} described at smaller $p_t$ in terms of string fragmentation~\cite{lund}. Angular correlations of fragments from hard-scattered partons (jets) were observed on $(\eta,\phi)$ at larger $p_t$ and with increasing $\sqrt{s}$~\cite{jets}. The {\em two-component} model of p-p collisions, with string and parton fragments forming distinguishable soft and hard components, describes particle multiplicity distributions at large $\sqrt{s}$~\cite{2compold} and has also been introduced to the heavy-ion context~\cite{2compnew}. 

Two issues arise in a conventional study of parton fragment distributions: 1) the distribution of fragment momenta along the jet thrust axis (fragmentation function) and 2) the angular distribution of fragments relative to the thrust axis. The thrust axis (estimating the parton direction) may be inferred from a collection of hadrons (jet) identified with the scattered parton, or (especially in heavy ion collisions) a `high-$p_t$ leading particle' may be used to estimate the parton momentum direction and magnitude. In the present analysis we adopt no {\em a priori} jet or factorization hypothesis. We study minimum-bias two-particle distributions on transverse rapidity space $(y_{t1},y_{t2})$ to obtain fragment {\em distributions} (not fragmentation {\em functions}) and on angle space $(\eta_1,\eta_2,\phi_1,\phi_2)$ to obtain fragment angular correlations. Particles in pairs are treated symmetrically, as opposed to asymmetric `trigger' and `associated' particle combinations.

We observe that hard-component correlations obtained with this minimum-bias analysis, in contrast to jet correlations obtained with the conventional leading- or trigger-particle approach, represent the {\em majority} of parton fragment pairs, those with $p_{t1} \sim p_{t2} \sim 1$ GeV/c. Analysis of nonperturbative low-$Q^2$ parton fragment correlations has motivated some novel techniques, including use of transverse rapidity $y_t$ rather than momentum $p_t$ and formation of {\em joint} (2D) angular {\em autocorrelations}. Symmetric analysis of fragment pairs requires a generalized treatment of fragmentation functions and jet angular correlations. An immediate consequence has been observation of substantially asymmetric angular correlations about the thrust axis for low-$Q^2$ parton collisions. While we attempt to understand QCD in A-A collisions we should also revisit its manifestations in elementary collisions, where novel phenomena are still emerging.

\section{String and low-$Q^2$ parton fragment correlations on $(y_{t1},y_{t2})$ and $(\eta_\Delta,\phi_\Delta)$}

\begin{figure}[h]
\begin{minipage}{18pc}
\begin{center}
 \includegraphics[width=8.5pc,height=8pc]{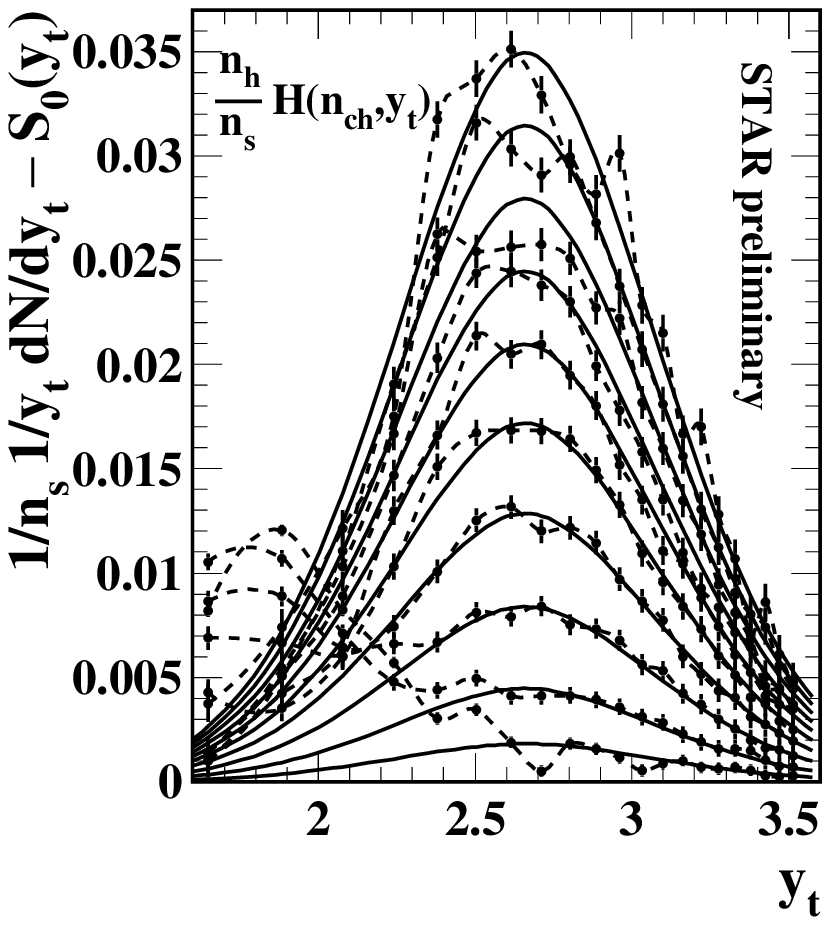} 
\includegraphics[width=8.5pc,height=8pc]{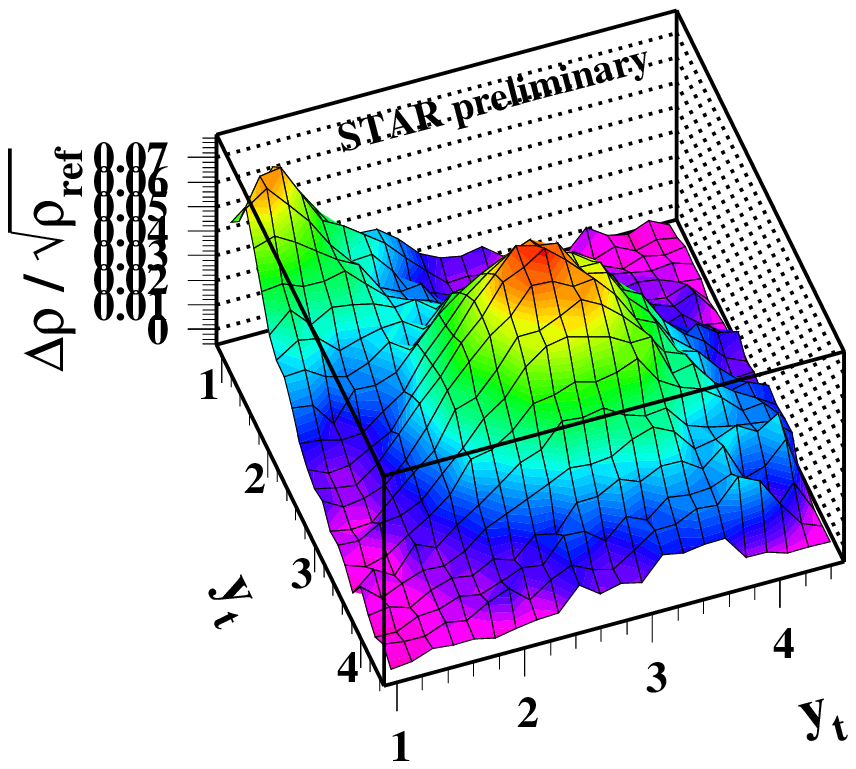}
\end{center} 
\end{minipage}
\hfil
\begin{minipage}{20pc}
\begin{center}
 \includegraphics[width=9.5pc]{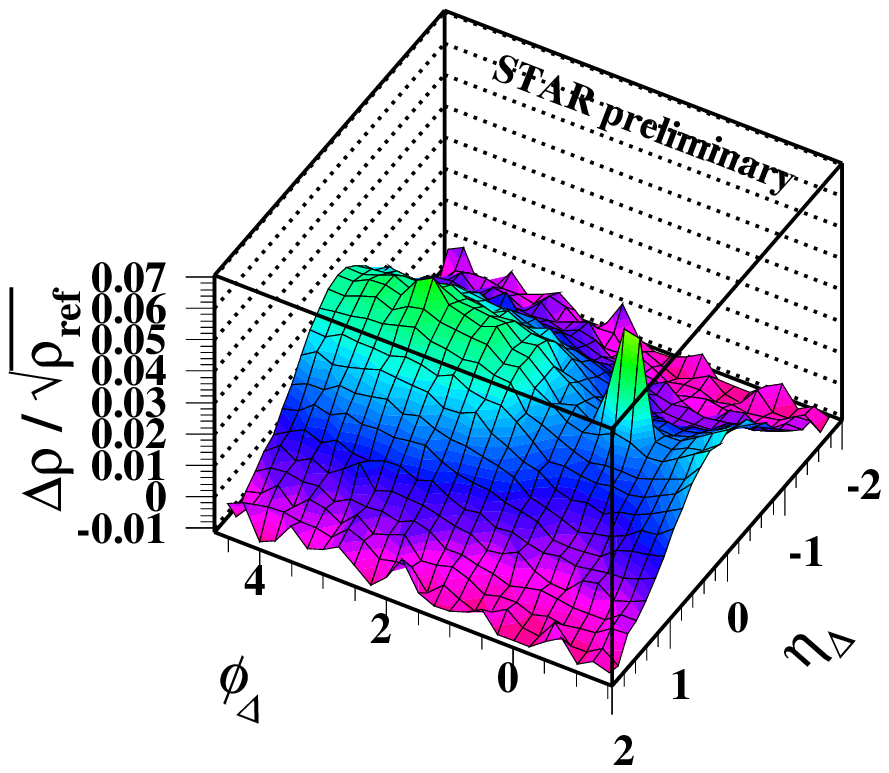} 
\includegraphics[width=9.5pc]{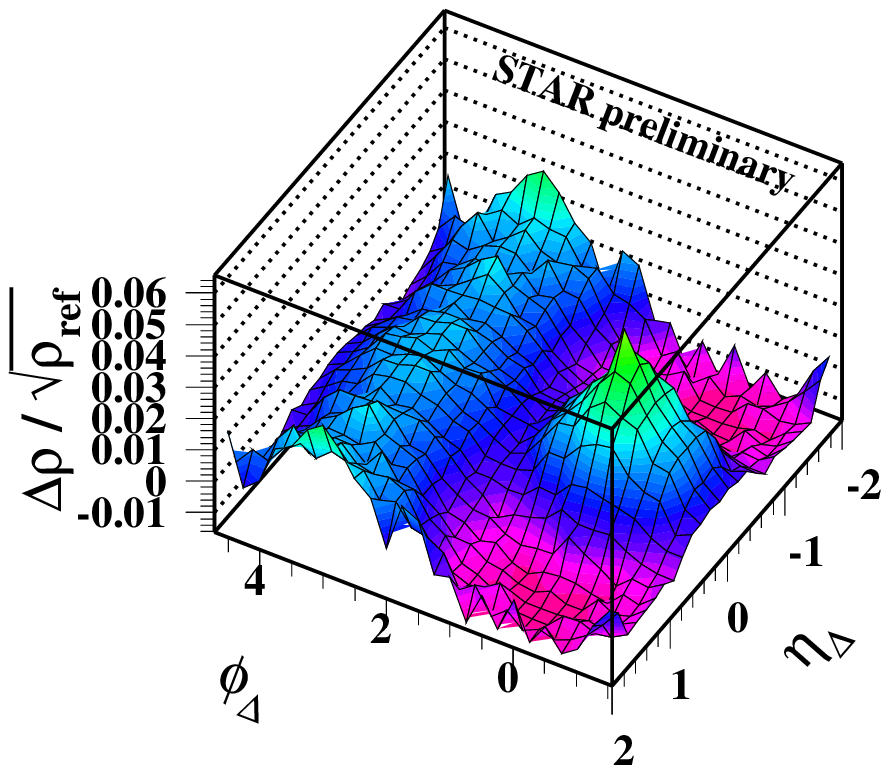} 
\end{center}
\end{minipage}\hspace{0pc}%
\caption{\label{lowq2} Hard components of p-p $p_t$ spectra {\em vs} $n_{ch}$ plotted on transverse rapidity $y_t$; low-$Q^2$ parton and string fragment distributions on $(y_{t1},y_{t2})$ and angular difference variables $(\eta_\Delta,\phi_\Delta)$.}
\end{figure}

In a study of $p_t$ spectra from p-p collisions at $\sqrt{s} = 200$ GeV we observed that the spectra can be separated into a soft component (string fragments - longitudinal fragmentation) described by a L\'evy distribution on transverse mass $m_t$ and a `semi-hard' component (parton fragments - transverse fragmentation) described by a gaussian distribution on transverse rapidity $y_t$, based on systematic variation with event multiplicity $n_{ch}$.
The hard component is interpreted as fragments of {\em minimum-bias} (mainly low-$Q^2$) partons. When a soft-component L\'evy distribution (defined as the limiting case of the $n_{ch}$ dependence) is subtracted from $p_t$ spectra for ten multiplicity classes we obtain distributions in Fig.~\ref{lowq2} (first panel), described by hard-component model distributions (solid curves) with gaussian shape essentially independent of $n_{ch}$. We have transformed to transverse rapidity $y_t \equiv \ln\{(m_t + p_t)/m_0\}$ with $m_0$ assigned as the pion mass to provide a `native' description of jets as hadron fragments from a moving source. 

That novel single-particle result motivated a follow-up study of two-particle correlations on transverse rapidity~\cite{jeffismd}. The minimum-bias distribution in Fig.~\ref{lowq2} (second panel) represents all event multiplicities and charge combinations within the STAR $(\eta,\phi)$ detector acceptance. Separate soft and hard components are evident, as is the correspondence with hard-component gaussians in the first panel. The $(y_{t1},y_{t2})$ space can thus be used as a cut space to study trends of corresponding angular correlations on pseudorapidity $\eta$ and azimuth $\phi$ for soft (string fragment) and hard (parton fragment) components, as shown in the third and fourth panels of Fig.~\ref{lowq2}.

A sequence of several analysis steps has led from study of single-particle $p_t$ spectra in a two-component context to isolation of soft and hard spectrum components to corresponding two-particle correlations on transverse rapidity to jet-like angular correlations obtained without imposition of a jet hypothesis. We thus achieve a model-independent analysis of parton scattering and fragmentation. Angular {\em autocorrelations} are not dependent on a leading or trigger particle and provide unprecedented access to low-$Q^2$ partons. The lower limit on fragment $p_t$ is determined by the fragmentation process itself, not the analysis method. We observe jet correlations in p-p collisions down to $p_t$ = 0.35 GeV/c for both particles. 
The main subject of this paper is the properties of fragment distributions on rapidity and angle from low-$Q^2$ partons in p-p collisions.

\section{Analysis method and data}


Correlation measure $\Delta \rho / \sqrt{\rho_{ref}}$ is closely related to Pearson's normalized covariance or {\em correlation coefficient}~\cite{pearson}. For event-wise particle counts $n_a$ and $n_b$ in histogram bins $a$ and $b$ on single-particle space $x$ and pair counts in corresponding bin $(a,b)$ on two-particle space $(x_1,x_2)$ we obtain Pearson's coefficient $ r_{ab} \equiv \overline{(n - \bar n)_a(n - \bar n)_b} / \sqrt{\overline{(n - \bar n)_a^2}\, \,\overline{(n - \bar n)_b^2}}$ averaged over an event ensemble. Pearson's coefficient is approximated by density ratio $\Delta \rho / \sqrt{\rho_{ref}} \equiv 1/\epsilon_x \,\, \overline{(n - \bar n)_a(n - \bar n)_{b}} / \sqrt{\bar n_a\, \bar n_{b}}$, where $\epsilon_x$ is the histogram bin size on $x$ and Poisson values of the number variances in the denominator have been substituted. $\Delta \rho / \sqrt{\rho_{ref}}$, estimating the density of correlated pairs {\em per particle}, is our basic correlation measure for two-particle distributions on $(y_{t1},y_{t2})$, $(\eta_{1},\eta_{2})$ and $(\phi_{1},\phi_{2})$. To view angular correlations compactly we combine spaces $(\eta_{1},\eta_{2})$ and $(\phi_{1},\phi_{2})$ by means of a {\em joint autocorrelation}, providing projection ({\em by averaging}) to a lower-dimensional space with essentially no information loss. An autocorrelation for 1D primary space $x$ is constructed by averaging coefficients $r_{a,a+k}$ over index $a$ along the $k^{th}$ diagonal on $(x_1,x_2)$. For angle space $(\eta,\phi)$ we average simultaneously along diagonals on $(\eta_{1},\eta_{2})$ and $(\phi_{1},\phi_{2})$ to obtain a {\em joint} autocorrelation, a projection by averaging of the full two-particle angle space onto its {\em difference variables} $\eta_\Delta \equiv \eta_1 - \eta_2$ and $\phi_\Delta \equiv \phi_1 - \phi_2$.


Data for this analysis were selected from 12 million p-p minimum-bias collision events at $\sqrt{s_{NN}} = 200$ GeV obtained with the STAR detector at RHIC~\cite{starnim}. Non-single-diffractive collisions were defined by a trigger coincidence between beam-beam counters positioned symmetrically in $3.5 \leq |\eta| \leq 5.0 $~\cite{starnim}. Particle tracks were reconstructed with the STAR Time Projection Chamber (TPC) operating within a uniform 0.5~T magnetic field parallel to the beam axis. Tracks were accepted within full azimuth, $ |\eta| \leq 1.0$ and $0.15 \leq p_{t} \leq 6$~GeV/c. TPC tracks and transverse beam position were used to reconstruct the primary or collision vertex. 



\section{p-p number correlations on $(y_{t1},y_{t2})$} \label{ytytsect}

\begin{figure}[h]
\begin{minipage}{18pc}
\begin{center}
 \includegraphics[width=8.5pc,height=8pc]{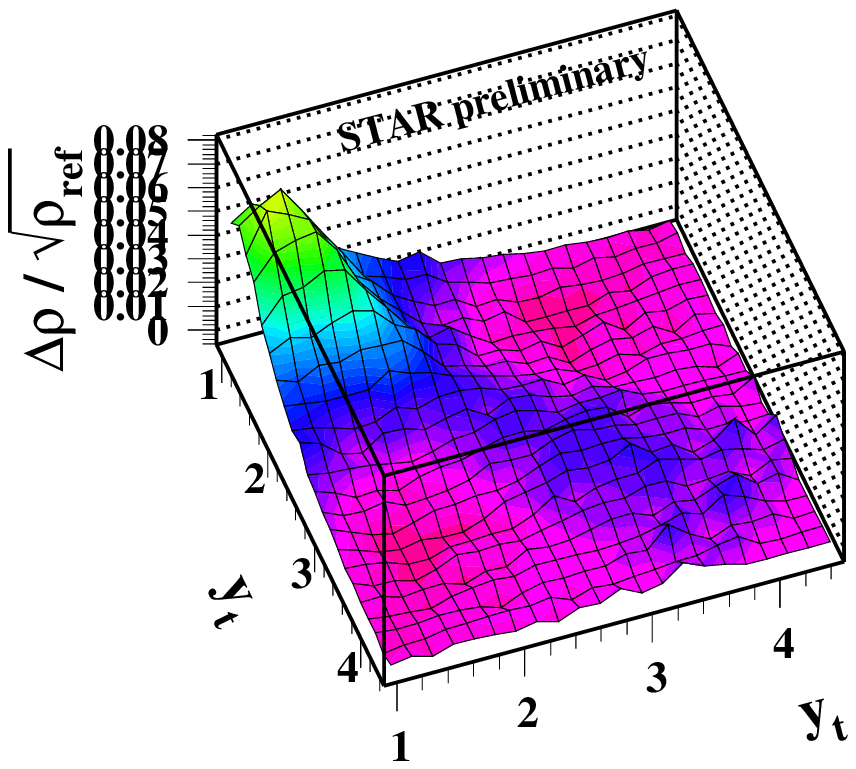} 
\includegraphics[width=8.5pc,height=8pc]{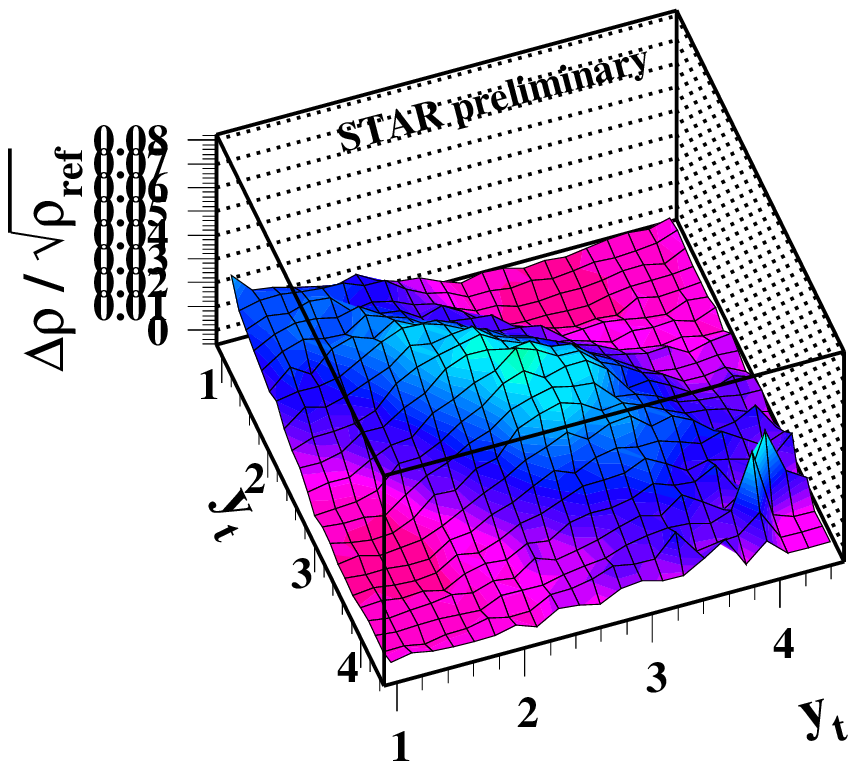}
\end{center} 
\end{minipage}
\hfil
\begin{minipage}{20pc}
\begin{center}
 \includegraphics[width=9.5pc]{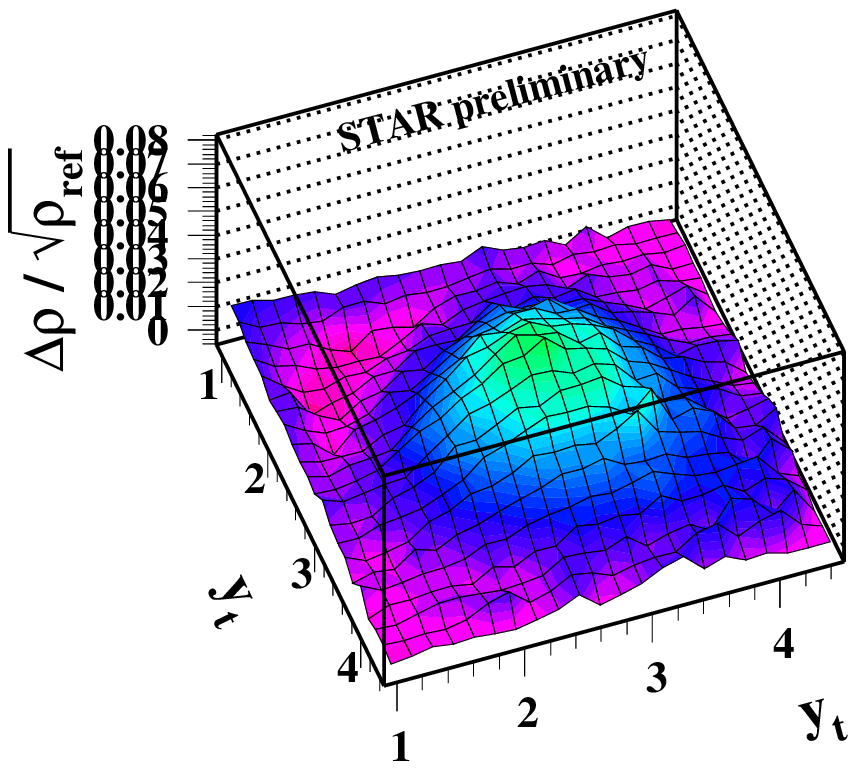} 
\includegraphics[width=9.5pc]{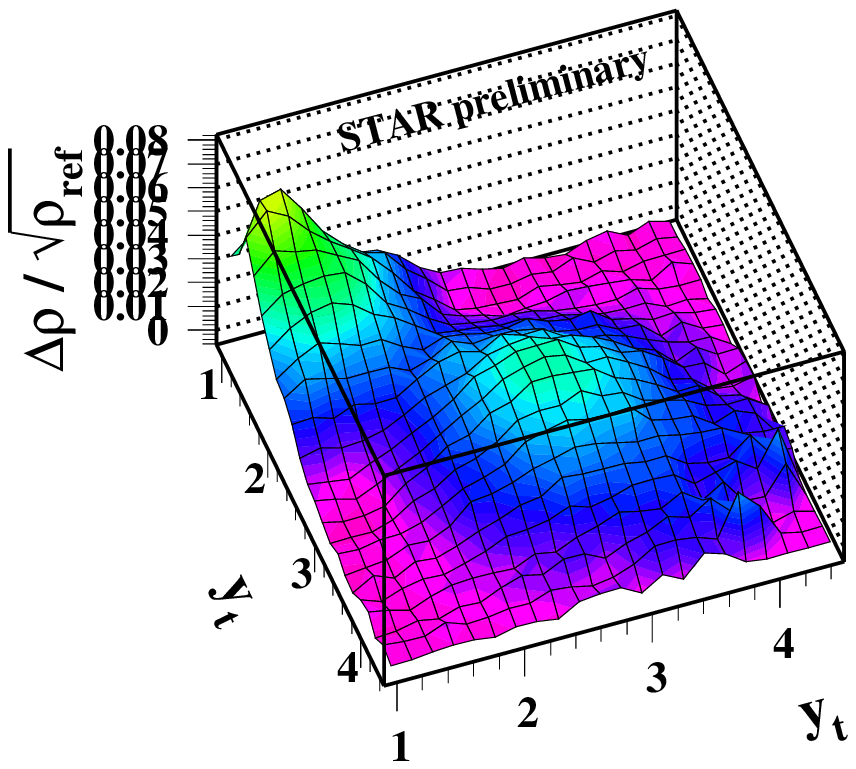} 
\end{center}
\end{minipage}\hspace{0pc}%
\caption{\label{ytyt} Number correlations on transverse rapidity $(y_t,y_t)$ from string and parton fragments for same-side (left) and away-side (right) pairs, and for like-sign and unlike-sign combinations.}
\end{figure}

Use of transverse rapidity $y_t$ to study parton fragmentation insures more compact distributions and relates directly to fundamental QCD aspects such as the double log approximation. It also provides a common basis, with longitudinal rapidity, for comparing string and parton fragmentation. Particle pairs from nuclear collisions can be separated on azimuth difference variable $\phi_\Delta \equiv \phi_1 - \phi_2$ into  {\em same-side} (SS)  ($|\phi_\Delta| < \pi/2$) and {\em away-side} (AS) ($|\phi_\Delta| > \pi/2$) pairs. Fig.~\ref{ytyt} shows same-side (left two panels) and away-side (right two panels) correlations of the form $\Delta \rho / \sqrt{\rho_{ref}}$ (normalized covariance) distributed on transverse rapidity subspace $(y_{t1},y_{t2})$. Each pair of panels represents like-sign (LS) and unlike-sign (US) charge combinations (left and right respectively). Structure can be separated into a soft component ($y_t < 2$ for both pair partners) and a hard component ($y_t > 2$ for both partners). Both components are strongly dependent on charge-sign combination and azimuth opening angle (same-side or away-side pairs). Description of the correlation structure is simplest in terms of sum and difference variables $y_{t\Sigma} \equiv y_{t1} + y_{t2}$ and $y_{t\Delta} \equiv y_{t1} - y_{t2}$.  

In the left (SS) panels, the LS soft component is interpreted as quantum correlations (HBT). The LS hard component along the diagonal is small and may itself be dominated by quantum correlations (from parton fragmentation). The US hard component is a peak at $y_t \sim 2.8$ ($p_t \sim$ 1 GeV/c) elongated along $y_{t\Sigma}$. The US hard component runs continuously into the US soft component at lower $y_t$, which is suppressed due to transverse momentum conservation. 
In the right (AS) panels, the hard-component peaks for LS and US pairs are nearly symmetric about their centers and much broader on $y_{t\Delta}$ than their SS US counterpart, with rapid falloff below $y_{t\Sigma} \sim 4$ (hadron $p_t \sim 0.5$ GeV/c) and nearly equal amplitudes. The large US soft component represents longitudinal string fragmentation, subject to local momentum and charge conservation (charge ordering on $z$)~\cite{lund} and therefore contributing negligibly to AS LS correlations.


\section{p-p number correlations on $(\eta_\Delta,\phi_\Delta)$}

\begin{figure}[h]
\begin{minipage}{18pc}
\begin{center}
 \includegraphics[width=8.5pc,height=8pc]{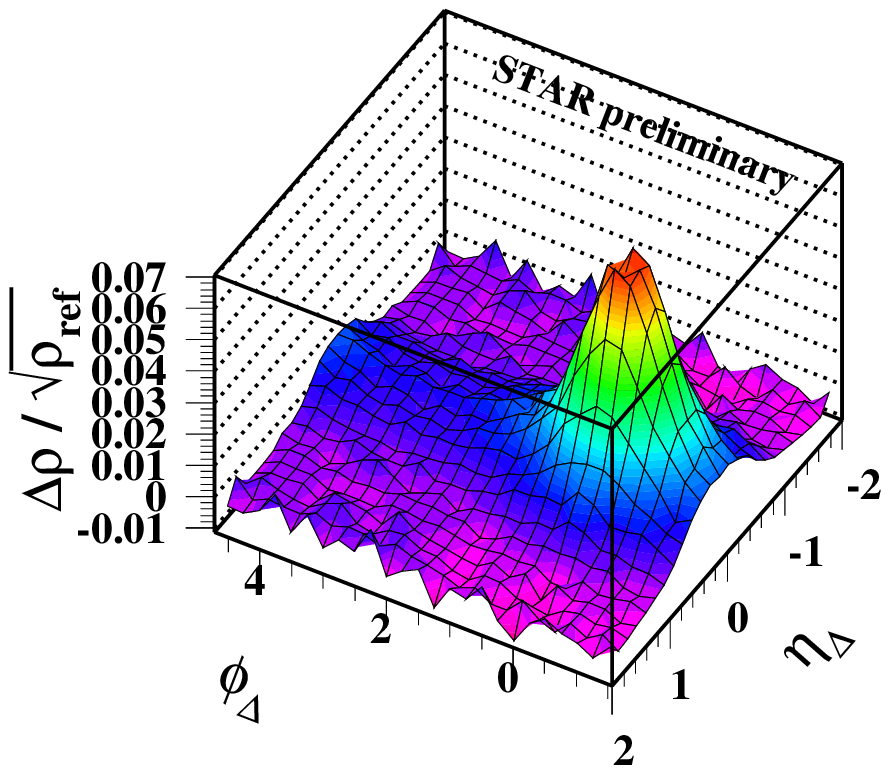} 
\includegraphics[width=8.5pc,height=8pc]{detadphidata6xyz}
\end{center} 
\end{minipage}
\hfil
\begin{minipage}{20pc}
\begin{center}
 \includegraphics[width=9.5pc]{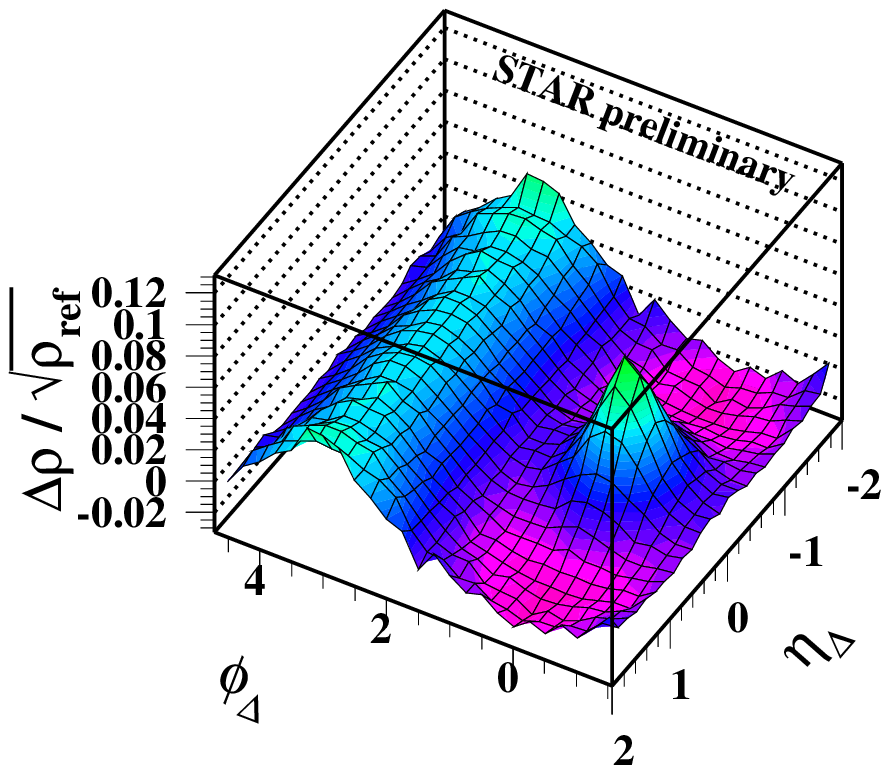} 
\includegraphics[width=9.5pc]{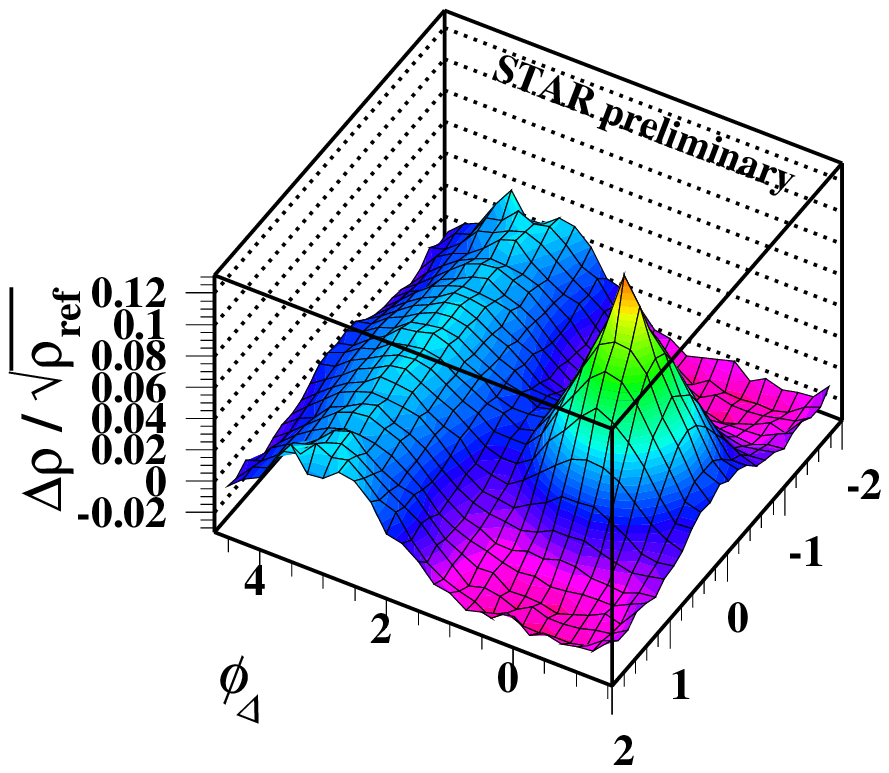} 
\end{center}
\end{minipage}\hspace{0pc}%
\caption{\label{etaphi}  Number correlations on $(\eta_\Delta,\phi_\Delta)$ for soft-component (left panels) and hard-component (right panels) pairs, and for like-sign and unlike-sign charge combinations.}
\end{figure}

The $(y_{t1},y_{t2})$ correlations in the previous section are directly related to angular correlations on $(\eta,\phi)$. To isolate soft and hard components of p-p angular correlations we define soft pairs by $y_t < 2$ ($p_t < 0.5$ GeV/c) and hard pairs by $y_t > 2$ for each particle. Fig.~\ref{etaphi} shows minimum-bias correlations (all pairs for all multiplicity classes) of the form $\Delta \rho / \sqrt{\rho_{ref}}$ on $(\eta_\Delta,\phi_\Delta)$ for the soft component (left two panels) and hard component (right two panels), with charge combinations like-sign (LS) and unlike-sign (US) (left and right respectively). The first panel is dominated by a 2D gaussian peak at the origin representing quantum correlations. The US combination in the next panel is dominated by a 1D gaussian peak on $\eta_\Delta$ arising from local charge conservation during string fragmentation (producing charge ordering along the collision axis). That trend is suppressed near the origin (the depression on azimuth of the $\eta_\Delta$ gaussian) due to local transverse-momentum conservation. The narrow peak at the origin represents electron-positron pairs from photon conversions. Except for the HBT contribution all structure in the soft component is apparently a consequence of local measure conservation during string fragmentation. The resulting `canonical suppression' is dependent on local particle density (event multiplicity).

The hard-component correlations in the right panels consist of a same-side peak at the origin and an away-side ridge. The LS same-side peak may be due to quantum correlations rather than jet fragmentation {\em per se}. The US same-side peak represents angular correlations of parton fragments (jet cone). The away-side hard-component correlations for LS and US pairs are essentially identical in shape and amplitude and reflect momentum conservation between scattered partons (dijets), including lack of structure on $\eta_\Delta$ due to the broad distribution of parton-collision centers of momentum. These hard-component $(\eta,\phi)$ systematics, fully consistent with conventional expectations for high-$p_t$ jet angular correlations, are observed in this study for pairs of particles with {\em both} $p_t$s as low as 0.35 GeV/c ($y_t \sim 1.6$), {\em much lower than previously observed with leading-particle methods}. 
In what follows we emphasize low-$Q^2$ parton fragmentation.

\section{Parton fragmentation: linear {\em vs} logarithmic presentation}


We first consider parton fragments distributed on transverse momentum. A {\em fragmentation function} describes the {\em conditional} distribution of fragment momenta along a jet thrust axis given the parton momentum (as estimated directly from collision kinematics or from  the observed fragment distribution or jet). Hadron momentum is conventionally normalized by the parton momentum/energy ({\em e.g.,} $x_p = p_t/E_{jet}$), represented by $E_{jet}$ (reconstructed jets), $\sqrt{s}/2$ (e$^+$-e$^-$) or $Q/2$ (e-p deep-inelastic scattering or DIS). Alternatively, relative fragment momentum is measured logarithmically by  $\xi \equiv \ln(1/x)$. 


\begin{figure}[h]
\begin{minipage}{19pc}
\begin{center}
 \includegraphics[width=9pc,height=8.8pc]{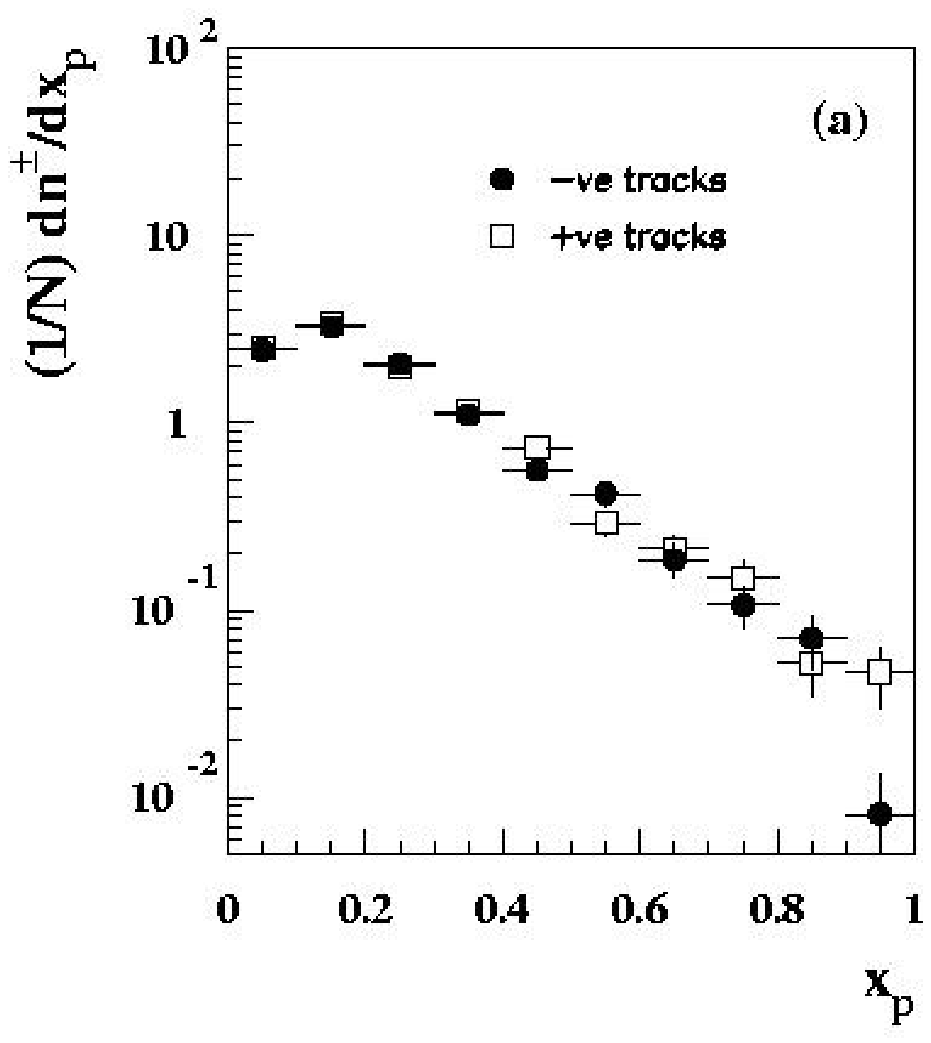} 
 \includegraphics[width=9pc,height=8.5pc]{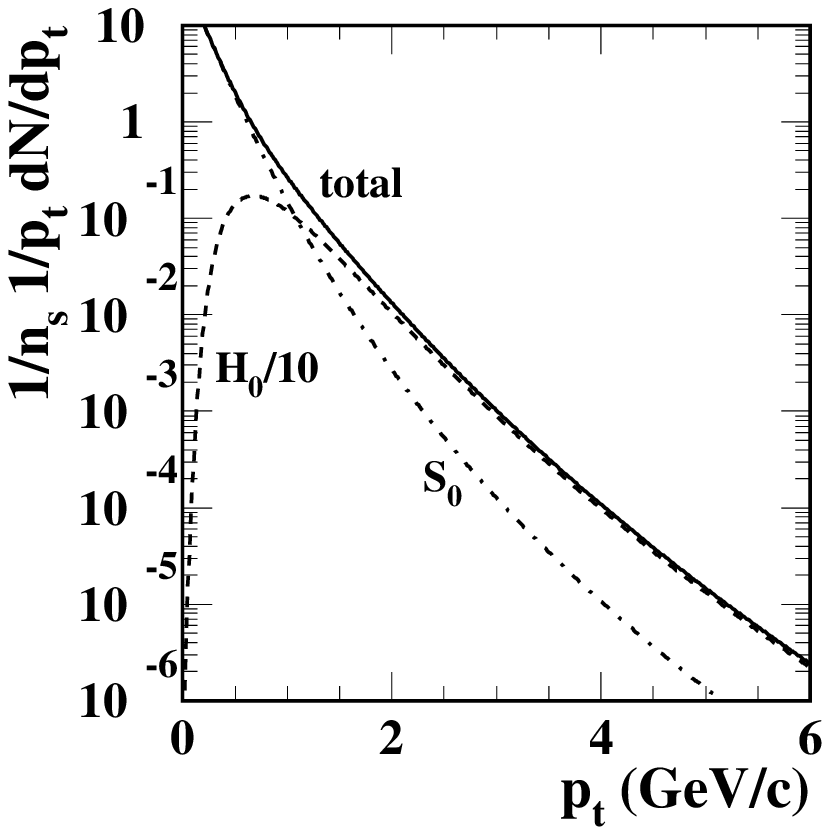} 
\end{center} 
\end{minipage}
\hfil
\begin{minipage}{18pc}
\begin{center}
 \includegraphics[width=8.5pc,height=8.5pc]{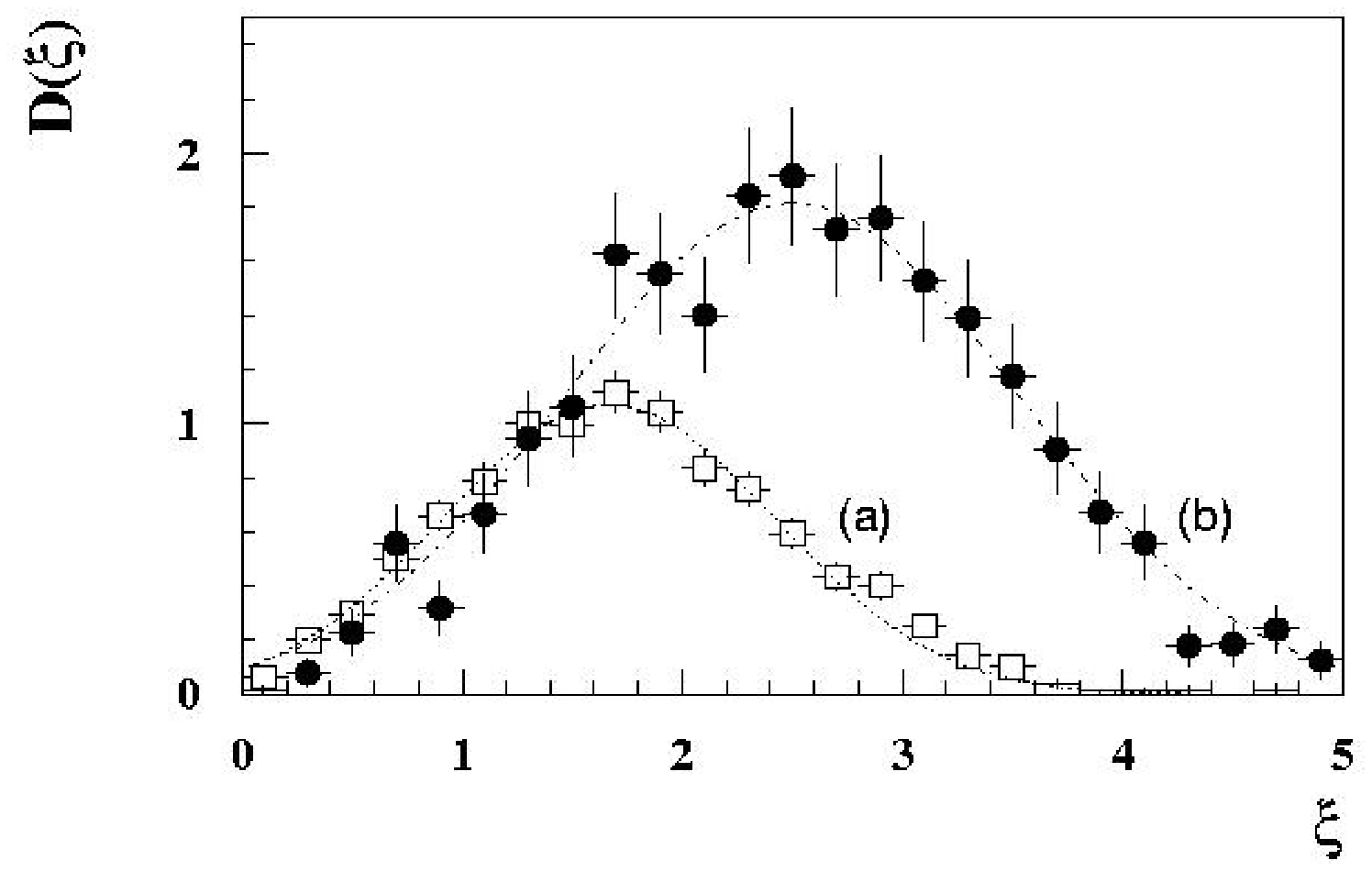} 
\includegraphics[width=8.5pc,height=8.5pc]{ppcomm12xyz} 
\end{center}
\end{minipage}\hspace{0pc}%
\caption{\label{linlog} DIS fragmentation functions on a semi-log format, $S_0$ (soft) and $H_0$ (hard) components of the p-p $p_t$ spectrum on a similar format, fragmentation functions on logarithmic momentum variable $\xi$ and p-p $p_t$ spectrum hard-component on transverse rapidity $y_t$.}
\end{figure} 
 
One semilog plotting convention is a logarithmic frequency distribution on linear transverse momentum, or normalized ratio $x_p = p_t / E_{jet}$, as shown in Fig.~\ref{linlog} (first panel) from e-p deep-inelastic scattering~\cite{h1}. Those distributions are similar to the hard component $H_0$ of the p-p single-particle $p_t$ spectrum when plotted on $p_t$ shown as the dashed curve in Fig.~\ref{linlog} (second panel). The fragmentation function plotted on linear $p_t$ is {\em approximately} exponential. An alternative convention is a linear frequency distribution on logarithmic transverse-momentum variable $\xi \equiv \ln(1/x_p) \simeq \ln (E_{jet} / p_t)$ as shown in Fig.~\ref{linlog} (third panel)~\cite{h1}. Those fragmentation functions (for two $Q^2$ intervals) can be compared with the hard components obtained from RHIC p-p spectra plotted on transverse rapidity in the fourth panel (the sense of increasing $p_t$ is reversed in the two cases). The conditional (on parton $Q/2$) fragmentation functions on $\xi$ in the third panel are approximately gaussian (further discussed below), as are the minimum-bias fragment distributions (no parton condition) in the fourth panel.

\section{Fragmentation functions on transverse rapidity}

\begin{figure}[h]
\begin{minipage}{19pc}
\begin{center}
 \includegraphics[width=19pc,height=8.5pc]{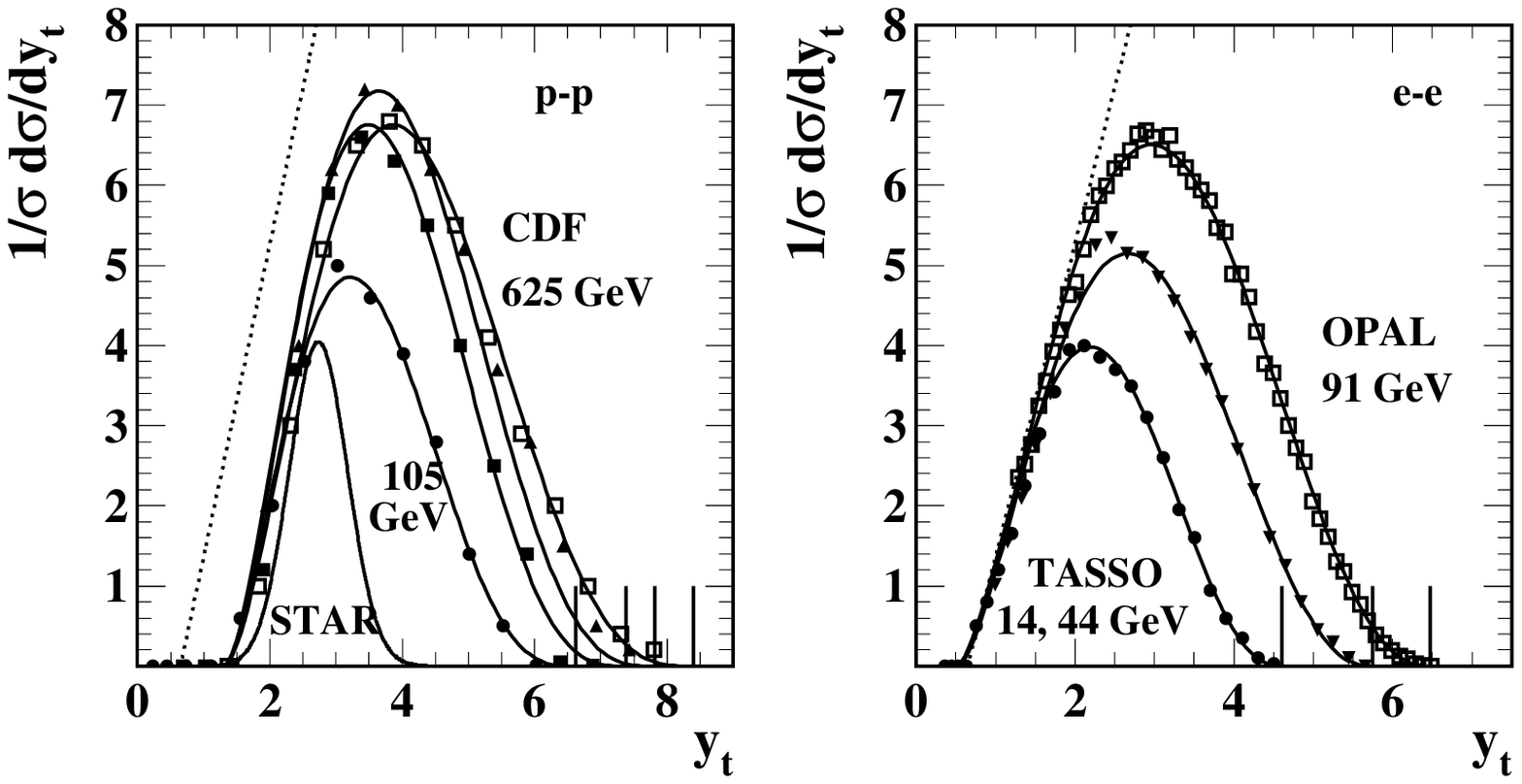} 
\end{center} 
\end{minipage}
\hfil
\begin{minipage}{19pc}
\begin{center}
\includegraphics[width=8.5pc,height=8.5pc]{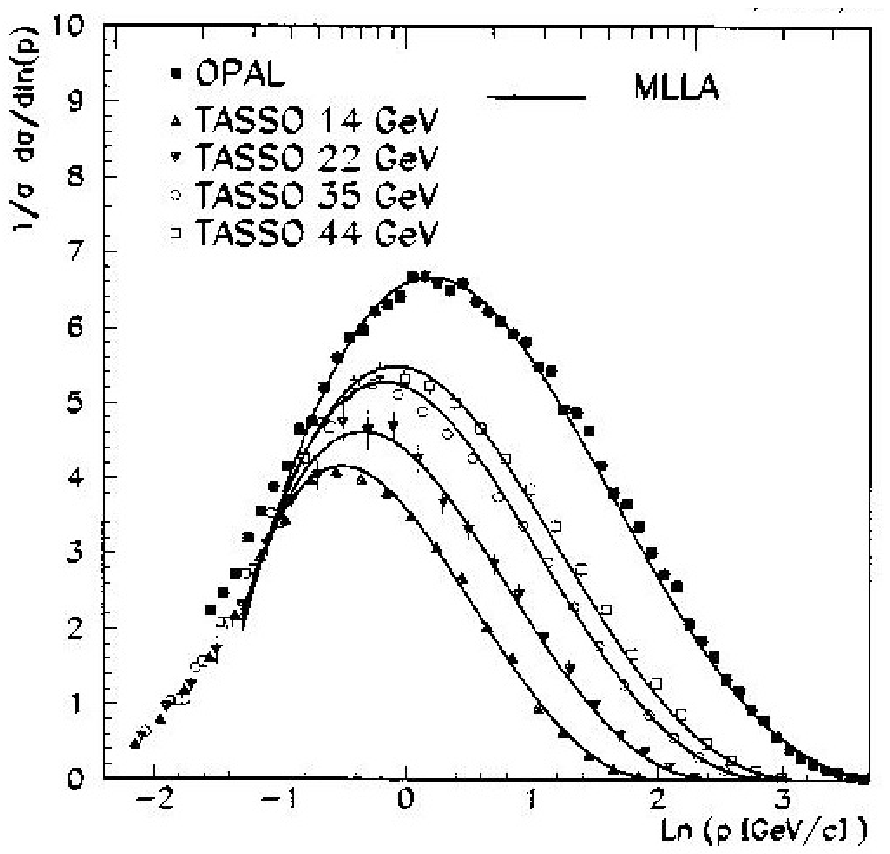} 
 \includegraphics[width=8.2pc,height=8.5pc]{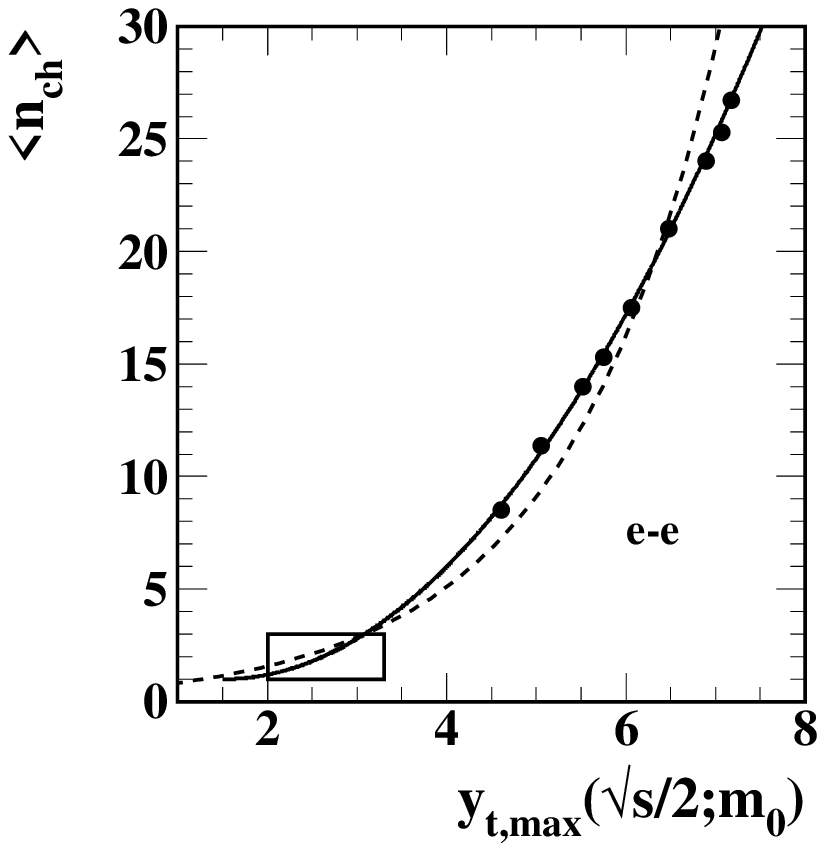} 
\end{center}
\end{minipage}\hspace{0pc}%
\caption{\label{fragonyt} Fragmentation functions from CDF, OPAL and TASSO plotted on transverse rapidity $y_t$ and momentum as $\ln(p)$; LEP jet fragment multiplicities plotted on parton rapidity $y_{t,max}$.
}
\end{figure}

The comparisons above motivated us to replot measured fragmentation functions on transverse rapidity $y_t$ as an {\em infrared safe} logarithmic momentum variable. In Fig.~\ref{fragonyt} (first two panels) we have replotted fragmentation functions from several collisions systems~\cite{cdf,opal} on transverse rapidity $y_t(p_t;m_0) \equiv \ln\{(\sqrt{p_t^2 + m_0^2} + p_t) / m_0 \}$, with $m_0$ assigned as the pion mass for all particles. We find that fragmentation functions so plotted are well described by beta distribution $\beta(x;p,q) = x^{p-1} (1-x)^{q-1} / B(p,q)$, with beta function $B(p,q) \equiv \Gamma(p+q)/\Gamma(p)\Gamma(q)$. The {\em standard} beta distribution defined on $x \in [0,1]$ is rescaled to $\beta(y_t;p,q)$ on $[y_{t,min},y_{t,max}]$. We observe that $y_{t,min}$ is 0.5 ($p_t \sim 0.075$ GeV/c) for e-e collisions and 1.5 ($p_t \sim 0.25$ GeV/c) for p-p collisions, the latter endpoint presumably reflecting the presence of the spectator parton system in the p-p case. The parton momentum is represented by $y_{t,max}(X;m_0) \equiv \ln\{(\sqrt{X^2 + m_0^2} + X) / m_0 \}$, with $X = E_{jet}$, $\sqrt{s}/2$ or $Q/2$ depending on the collision system. $y_{t,max}$ values for each fragmentation function in Fig.~\ref{fragonyt} are marked with short vertical lines. For e-e collisions the beta parameters are $p = 2.75$ and $q = 3.5$ for the cases examined; for p-p collisions the values are $p \sim 2.8$ and $q \sim 4.2$. The hard component from the STAR two-component analysis of p-p collisions is included in the first panel as the small gaussian curve (describing {\em minimum-bias} parton fragments with no $y_{t,max}$ condition). The energies in that panel are {\em dijet} energies $E_{jj} = 2\,E_{jet}$. The data in the second panel are also plotted on $\ln(p)$ (comparable to $\xi$ but with the sense of increasing $p_t$ reversed) in the third panel. The solid curves are MLLA predictions~\cite{cdf}, which reflect momentum conservation and QCD branching above the peaks but deviate strongly from data at lower momenta. In contrast, the beta distribution provides a precise description of the entire fragment distributions on transverse rapidity in the left panels.


The distribution systematics for e-e collisions in the second panel, interpreted in the context of the double-log approximation (DLA), imply that the e-e jet multiplicity trend can be described by $\langle n_{ch} \rangle = 1 + A\, (y_{t,max} - y_{t,min})^2$, whereas the modified leading-log approximation (MLLA) prediction is $ \exp\{ a\,\ln(Q/2\Lambda)\}$~\cite{cdf}. The fourth panel shows measured jet multiplicities {\em vs} jet energy ($y_{t,max}$) for e-e collisions (points)~\cite{bethke}, our parameterization with $A = 0.8$ (solid curve) and the MLLA prediction with $a$ chosen to provide the best fit (dashed curve). The hatched region corresponds to the low-$Q^2$ partons which dominate a minimum-bias p-p analysis. By a similar argument we can represent the most probable point (mode) as $y_{t,peak} = B\,(y_{t,max} - y_{t,min}) + y_{t,min}$, where for the beta distribution and e-e collisions $B \equiv (p-1)/(p+q-2) = 0.45$, whereas the MLLA gives $\xi_{peak} \sim \ln(Q/2\Lambda)$. Fragment distributions on transverse rapidity are consistent with conventional treatments ({\em e.g.,} $\xi$) but are `infrared safe' as $p_t \rightarrow 0$.

\section{Symmetrized two-particle fragment distributions}

The conventional asymmetric treatment of parton fragments in terms of trigger and associated particles cannot access the low-$Q^2$ partons of greatest interest to us. We therefore symmetrize the analysis of fragment correlations, first for two-particle fragment distributions on $y_t$ and then for angular correlations. The same-side unlike-sign (US) $(y_t,y_t)$ correlations in Sec.~\ref{ytytsect} can be interpreted as a two-particle jet fragment distribution which we now model: we combine what is known about single-particle fragmentation functions with expectations for two-particle correlations to sketch the parameterization of a two-particle fragment distribution. 

\begin{figure}[h]
\begin{minipage}{9pc}
\begin{center}
 \includegraphics[width=9pc,height=9pc]{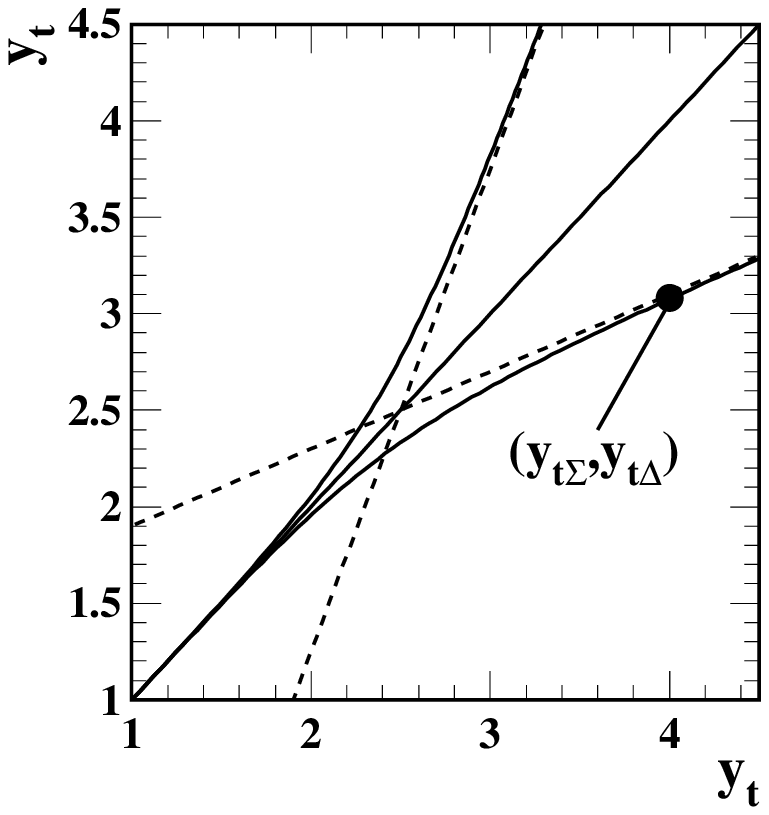} 
\end{center} 
\end{minipage}
\hfil
\begin{minipage}{19pc}
\begin{center}
 \includegraphics[width=18pc,height=9pc]{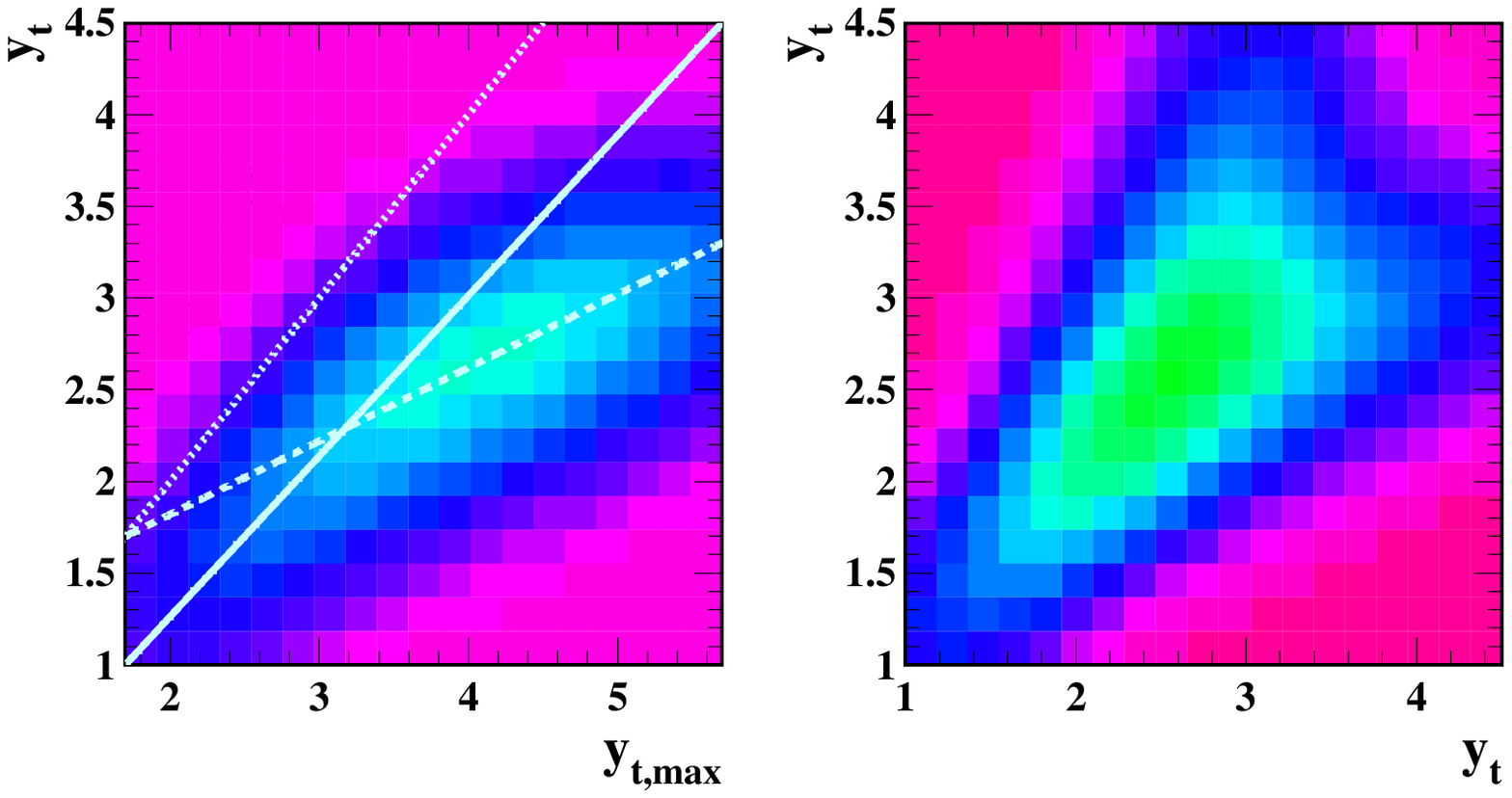} 

\end{center} 
\end{minipage}
\begin{minipage}{9pc}
\begin{center}
 \includegraphics[width=9pc,height=9pc]{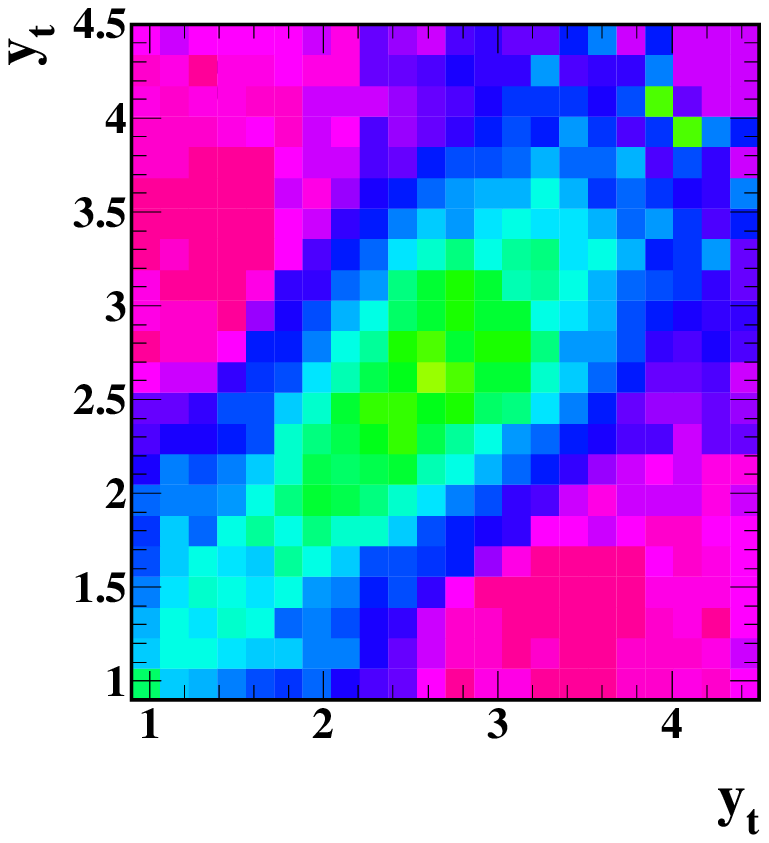} 
\end{center}
\end{minipage}\hspace{0pc}%
\caption{\label{consym} Locus of most-probable points for pair fragment distribution, fragment rapidity vs parton rapidity, symmetrized distribution, measured same-side unlike-sign fragment distribution.}
\end{figure}

In the previous section we presented single-particle fragmentation functions plotted on transverse rapidity $y_t$. An example of fragmentation functions plotted on $\xi$ from the H1 experiment at HERA is shown in Fig.~\ref{linlog} (third panel) for two intervals on $Q/2$. Based on those trends we sketch a 2D distribution relating parton momentum $Q/2$ and fragment momentum $p_t$, both in rapidity format, as shown in the first two panels of Fig.~\ref{consym}. The first panel shows the locus of modes (most probable points, dashed line with smaller slope) for a continuum of fragmentation functions varying with parton rapidity on the horizontal axis, with fragment rapidity on the vertical axis. To construct the two-particle distribution we require a curve smoothly joining that line of modes and the diagonal with unit slope (solid line) given by the solid curve. The plot is then symmetrized. In the second panel we sketch the corresponding 2D distribution of fragmentation functions {\em vs} parton rapidity $y_{t,max}$. The diagonal dotted line represents parton momentum $Q/2$ plotted as a rapidity on the fragment axis. The dashed line is the line of modes. The fragmentation function for a given parton momentum would be a vertical slice from  this distribution.

In the third panel we symmetrize the distribution from the second panel because in our $(y_t,y_t)$ analysis we correlate fragments with fragments symmetrically, rather than fragments with partons. That sketch can be compared with data in the fourth plot for unlike-sign same-side pairs -- the minimum-bias two-particle fragment distribution. For low-$Q^2$ partons, which dominate the minimum-bias parton distribution, the two-particle fragment distribution becomes symmetric about the sum diagonal (and the fragment number $\rightarrow 2$). This parton-fragment distribution is biased by the requirement of fragment {\em pairs}: `fragmentation' to single hadrons cannot contribute to this plot. Because of the symmetry, the more appropriate variables to describe the distribution are sum and difference rapidities $y_{t\Sigma} \equiv y_{t1}+y_{t2}$ and $y_{t\Delta} \equiv y_{t1}-y_{t2}$ as illustrated in the first panel. The comparison between sketch and data is informative. The small excess yield at low $y_t$ in the fourth panel is string fragments, partially suppressed by transverse momentum conservation.


\section{Jet morphology for low-$Q^2$ partons}


In the previous section we described parton fragment correlations on transverse rapidity. Jet structure is also characterized by two-particle angular correlations of hadron fragments, both intra-jet (within one jet) and inter-jet (between opposing dijets). The conventional method to describe angular correlations of fragments in the absence of full jet reconstruction is as a {\em conditional} distribution relative to a leading or trigger particle ({\em e.g.,} the highest-$p_t$ particle in an event) used as an estimator for the parton momentum. The distribution of `associated' particles relative to the trigger particle is an approximation to full jet reconstruction. Angular correlations relative to the trigger reveal 2D same-side peaks with (possibly different) widths on $\eta$ and $\phi$. From those widths the mean momenta of fragments transverse to the jet {\em thrust axis} in the two angular directions $(\eta,\phi)$ are inferred.

In this minimum-bias study we encounter fragment distributions with a most-probable $p_t$ of about 1 GeV/c. The most probable fragment multiplicity is 2, with approximately equal fragment momenta, and the parton $Q/2$ is thus somewhat more than 2 GeV---{\em comparable} to the intrinsic parton $k_t$. For jets from low-$Q^2$ partons we cannot differentiate trigger and associated particles. We want a symmetrized formulation of angular correlations relative to which the leading-particle conditional treatment is a {\em special case}. For {\em intra-jet} correlations we define a symmetrized expression relating angular correlation width and transverse momentum $j_t$ relative to thrust as
\bea \label{jt}
\overline{\langle j_{t\phi}^2 \rangle}_{12} \equiv \frac{\sqrt{\langle p^2_{t,1}\rangle \langle p^2_{t,2}\rangle}}{\sqrt{\langle p^2_{t,2}\rangle / \langle p^2_{t,1}\rangle} + \sqrt{\langle p^2_{t,1}  /  \rangle\langle p^2_{t,2}\rangle}}\left\{ \langle \sin^2 \Delta \phi \rangle_{SS} + 2 \langle \sin^2 \phi_1 \rangle\langle \sin^2  \phi_2 \rangle \right\} \\ \nonumber
\simeq \frac{(m_\pi / 2)^2\, \exp(y_{t\Sigma})}{2 \,\cosh(y_{t\Delta})}\left\{ \langle \sin^2  (\phi_\Delta / \sqrt{2}) \rangle_{SS} + 2 \left\langle \sin^2 \left(\phi_\Delta / 2 \sqrt{2}\right) \right\rangle^2  \right\}
\eea
for $\phi$, with a similar expression for $\eta$.

The inter-jet azimuth angular correlation of particle pairs from opposed jets (dijets) is wider than the intra-jet azimuth correlation. The excess width or `$k_t$ broadening' is attributed to the initial (intrinsic) transverse momenta $k_t$ of partons prior to scattering, which produces an {\em acoplanarity} of the two jets with respect to the collision axis. The symmetrized relation between angular correlation width and inferred $k_t$ is
\bea \label{kt}
\overline{\langle z^2 \rangle \, \langle k_{t\phi}^2 \rangle}_{12} \equiv \frac{\sqrt{\langle p^2_{t,1}\rangle\langle p^2_{t,2}\rangle}}{\sqrt{\langle p^2_{t,2}\rangle / \langle p^2_{t,1}\rangle} + \sqrt{\langle p^2_{t,1}  /  \rangle\langle p^2_{t,2}\rangle}} \left\{ \frac{ \langle \sin^2 \Delta \phi \rangle_{AS} -  \langle \sin^2 \Delta \phi \rangle_{SS}}{1-2\langle \sin^2 \Delta \phi \rangle_{SS}} \right\} \\ \nonumber
\simeq \frac{(m_\pi / 2)^2\, \exp(y_{t\Sigma})}{2 \,\cosh(y_{t\Delta})} \left\{ \frac{\langle \sin^2  (\phi_\Delta / \sqrt{2}) \rangle_{AS} - \langle \sin^2  (\phi_\Delta / \sqrt{2}) \rangle_{SS} }{1 - 2 \langle \sin^2  (\phi_\Delta / \sqrt{2}) \rangle_{SS}} \right\}.
\eea
Those symmetrized expressions are used to extract fragment transverse momenta relative to thrust axis from angular correlations from p-p collisions. Fragment angular correlations are thus obtained for previously inaccessible low-$Q^2$ partons.

\section{Angular correlation measurements} 

\begin{figure}[h]
\begin{minipage}{19pc}
\begin{center}
 \includegraphics[width=9pc,height=8.5pc]{etaphidata01} 
\includegraphics[width=9pc,height=8.5pc]{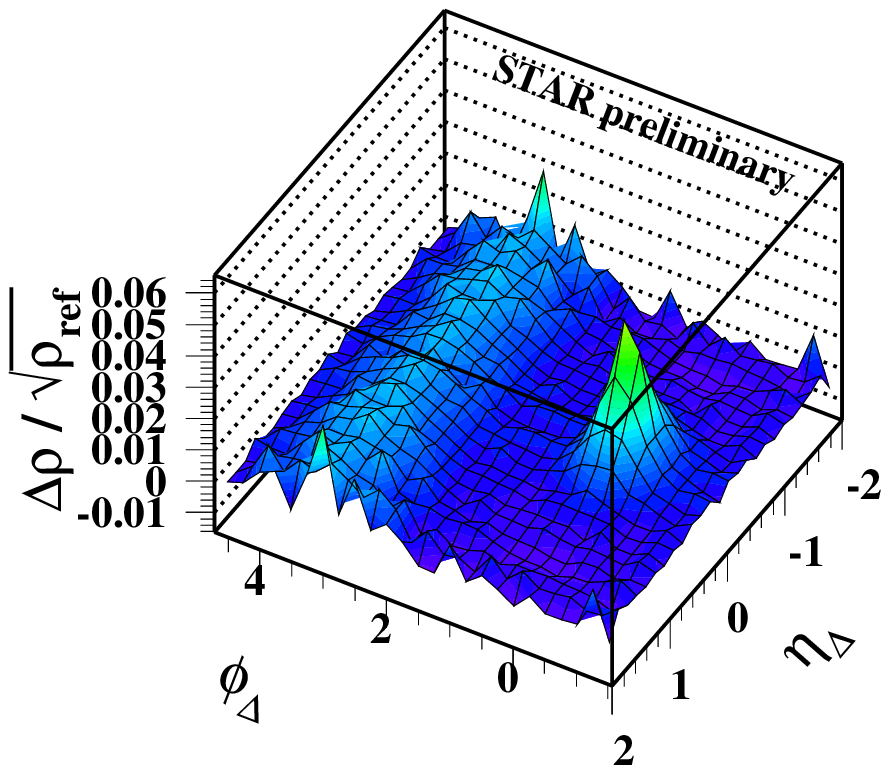}
\end{center} 
\end{minipage}
\hfil
\begin{minipage}{17pc}
\begin{center}
 \includegraphics[width=17pc]{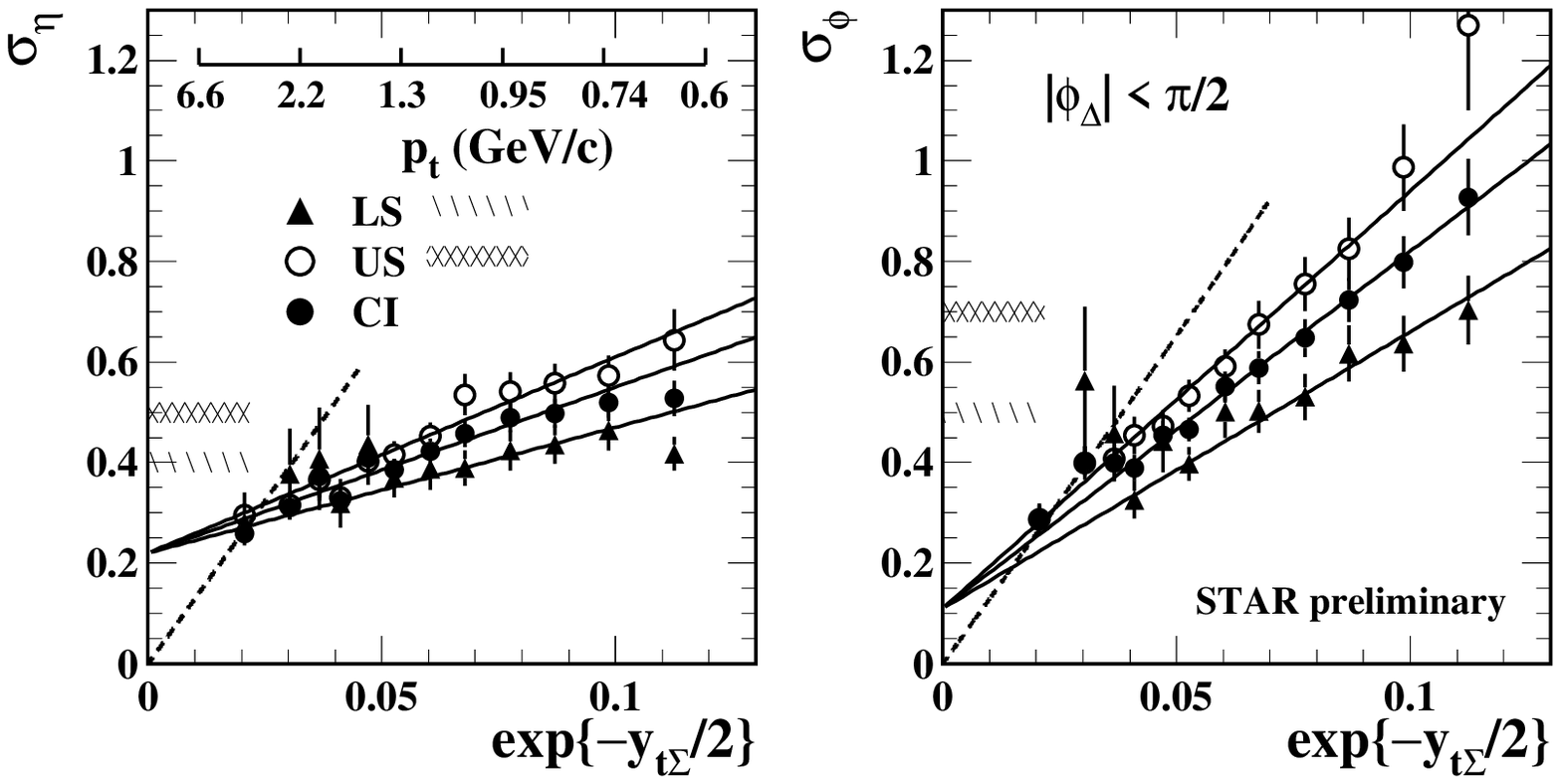} 
\end{center}
\end{minipage}\hspace{0pc}%
\caption{\label{angcorr}  Angular correlations for parton fragment pairs with smaller and larger geometric-mean $p_t$; measured angular widths of same-side peaks on $\eta$ and $\phi$ {\em vs} inverse mean $p_t$.}
\end{figure}

Fig.~\ref{jtkt} (first panel) shows two-particle correlations on transverse-rapidity space $(y_{t1},y_{t2})$ described on sum and difference variables $y_{t\Sigma} \equiv y_{t1} + y_{t2}$ and $y_{t\Delta} \equiv y_{t1} - y_{t2}$. Hard-component fractions for angular correlation measurements are defined by the grid of rectangles or bins (the narrow yellow lines) along $y_{t\Sigma}$ numbered $1, \cdots, 12$. The solid green boxes in the upper-right corner represent regions typically explored in {\em leading-particle} analyses based on high-$p_t$ `trigger' particles~\cite{leading}. The dashed extensions represent cuts for extended associated-particle conditions recently applied to heavy ion collisions~\cite{fuqiang}. The first two panels of Fig.~\ref{angcorr} show angular autocorrelations for the second and eleventh bins on $y_{t\Sigma}$ in Fig.~\ref{jtkt} as examples. In the first panel the most probable combination is two particles each with $p_t \sim 0.6$ GeV/c. The same-side peak (jet cone) is broad but well-defined. The away-side ridge (particle pairs from dijets) is uniform on $\eta_\Delta$ as expected. This jet correlation is a remarkable result for particles with such low $p_t$ and illustrates the power of the autocorrelation technique. The second panel shows correlations for the 11$^{th}$ bin, corresponding to $p_t \sim 2.5$ GeV/c for each particle. The same-side cone is much narrower, as is the away-side peak on azimuth $\phi$, a result more typical of a high-$p_t$ leading-particle analysis. 

The general trend is monotonic reduction of peak widths on  $(\eta_\Delta,\phi_\Delta)$ with increasing $y_{t\Sigma}$. The angular widths are determined by the mean fragment momentum $\hat \jmath_t$ perpendicular to the parton momentum (jet thrust) axis, relative to the particle $p_t$ in the form $\hat \jmath_t / p_t$.  We extract from those angular correlations near-side widths on $\eta_\Delta$ and $\phi_\Delta$, and an away-side width on $\phi_\Delta$. We fit the same-side ($|\phi_\Delta| < \pi/2$) angular correlations (jet cone) with a 2D Gaussian and extract widths on pseudorapidity and azimuth for like-sign (LS), unlike-sign (US) and total (CI) pairs. Those widths are plotted in the last two panels of Fig.~\ref{angcorr} {\em vs} $\exp\{-y_{t\Sigma} / 2\} \propto 1/\sqrt{p_{t1}\, p_{t2}}$, that is, proportional to the inverse geometric mean of the pair momenta. The corresponding mean particle $p_t$ is given at the top of the third panel. 

The dashed diagonal lines in those panels represent `$\hat \jmath_t$ scaling': the assumption that $\hat \jmath_t$ ({\em r.m.s.} $j_t$) is independent of hadron momentum as a limiting case for larger $p_t$. We observe a dramatic departure from that expectation for these low fragment momenta ($p_t < 2.5$ GeV/c). However, with the autocorrelation technique we observe that jet structure is very well defined, even down to hadron pairs with particle $p_t \sim 0.35$ GeV/c. The hatched regions to the left in those panels show the values measured for minimum-bias hard-component hadrons, which are weighted toward lower $p_t$. In  the following section we examine the systematics of $\hat \jmath_t$ {\em vs}  $y_{t\Sigma}$ separately on $\eta$ and $\phi$, and for values of parton $Q^2$ previously inaccessible.

\section{Inferred $\hat \jmath_t$ and $\hat k_t$ distributions}

\begin{figure}[h]
\begin{minipage}{9pc}
\begin{center}
 \includegraphics[width=8.5pc,height=8.5pc]{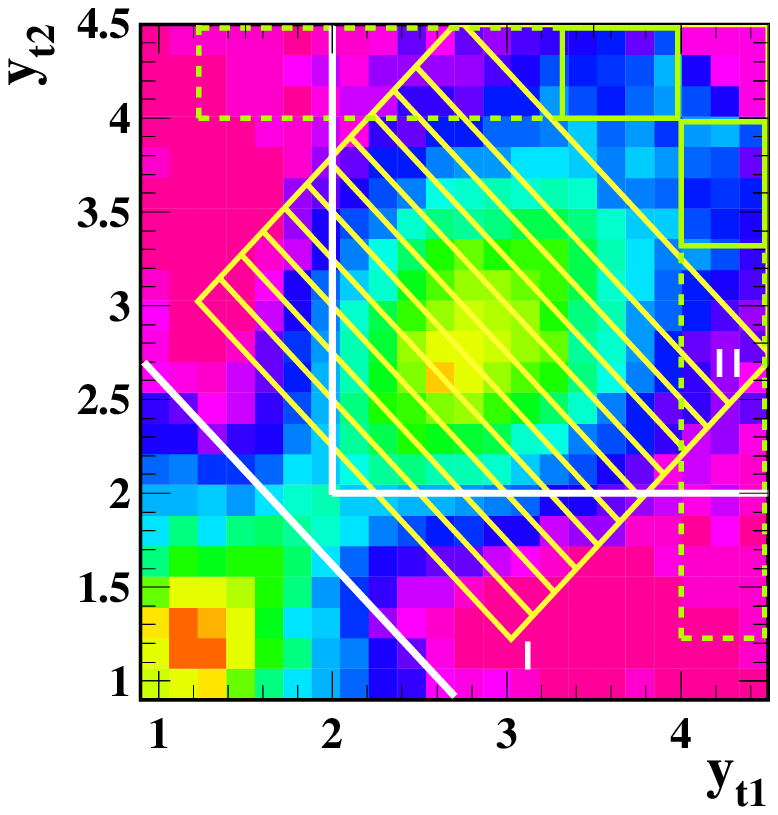} 
\end{center} 
\end{minipage}
\hfil
\begin{minipage}{19pc}
\begin{center}
 \includegraphics[width=18pc,height=8.5pc]{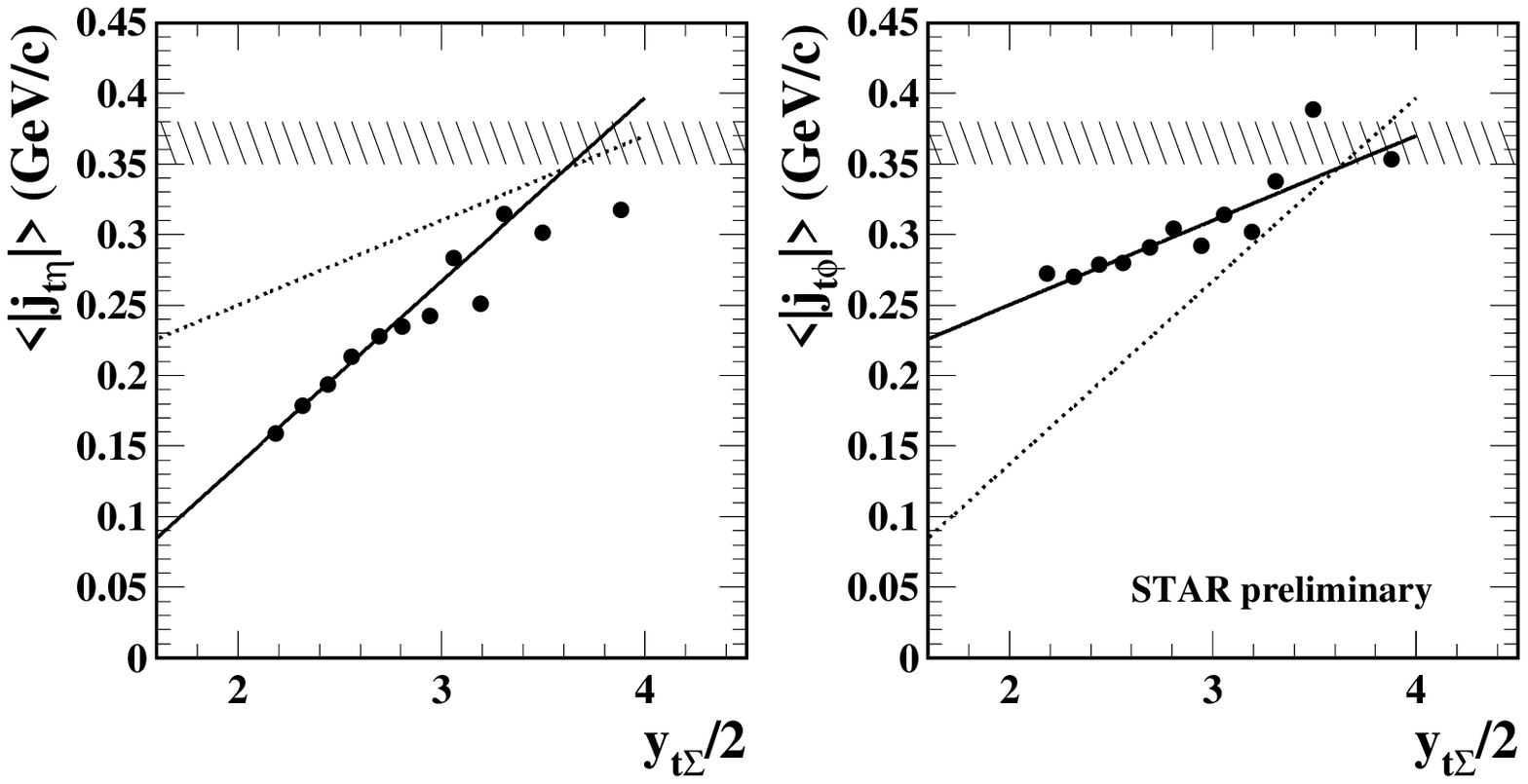} 
\end{center}
\end{minipage}\hspace{0pc}%
\begin{minipage}{9pc}
\begin{center}
 \includegraphics[width=8.5pc,height=8.5pc]{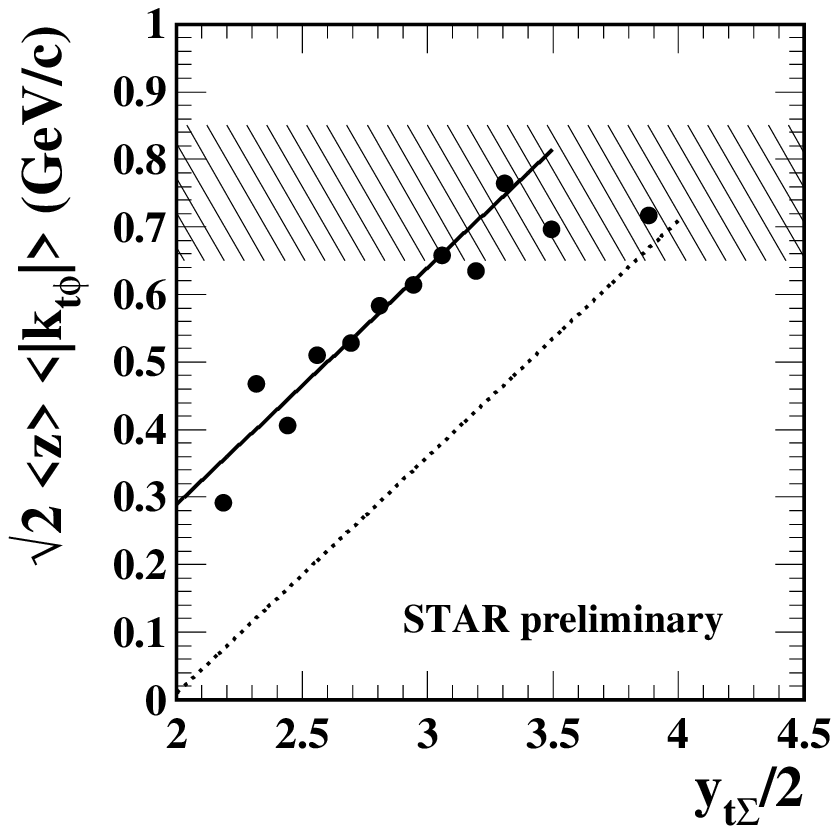} 
\end{center} 
\end{minipage}
\caption{\label{jtkt} Minimum-bias fragment distribution on $(y_{t1},y_{t2})$ showing cut grid along the diagonal used for the angular correlation study; $\hat \jmath_t$ and $\hat k_t$ values inferred from angular correlation study.}
\end{figure}

By applying Eq.~(\ref{jt}) 
with the values of $(y_{t\Sigma},y_{t\Delta})$ defined on the $(y_{t1},y_{t2})$ cut space as in Fig.~\ref{consym} (first panel) we obtain $\hat \jmath_t$ values for $x = \eta, \,\phi$ from the corresponding angular widths. In Fig.~\ref{jtkt} (second and third panels) we observe a strong decrease in both $\hat \jmath_t$ components with decreasing $y_{t\Sigma}$ ($\sim$ parton $Q^2$). We also observe substantial differences in the $\hat \jmath_t$ $\eta$ and $\phi$ components, converging however to the `$\hat \jmath_t$ scaling' expectation at larger $y_{t\Sigma}$ (hatched bands). 
It is notable that the orientation of the $\hat \jmath_t$ asymmetry (major axis lying in the $(\phi,p_t)$ plane) is parallel to the `contact plane' of the parton collision. That asymmetry orientation is orthogonal to what is observed for low-$Q^2$ jet fragments in central heavy ion collisions~\cite{axialci}.

We also measured the angular width of the {\em away-side} azimuth correlation of hadron pairs ($|\phi_1 - \phi_2| > \pi/2$) and compared it with the same-side azimuth correlation to infer the mean transverse momentum $\hat k_t$ of partons prior to scattering. 
We converted the angular widths to $\hat k_t$ values according to Eq.~(\ref{kt}),
where $z$ estimates the mean value of the ratio of `trigger particle' momentum to parton momentum. At larger $p_t$ that value is approximately 0.75, but is a falling function of $p_t$ for smaller fragment $p_t$, as in this case. Below $y_{t\Sigma}/2 \sim 3.3$ (hadron $p_t \sim 2$ GeV/c) we observe that $\hat k_t$ values fall below the $k_t$-scaling trend (hatched band) by an increasing amount with smaller $y_{t\Sigma}$ (smaller $Q^2$). The trend of Pythia $\hat k_t$ values (dotted line) is substantially displaced below the data.




\section{Summary}

We have presented a broad survey of newly-obtained two-particle correlations from 200 GeV p-p collisions at RHIC. Correlations from longitudinal string fragmentation and transverse scattered parton fragmentation are clearly distinguished. The former is dominated by local momentum and charge conservation. The latter provides new access to scattering and fragmentation of {\em minimum-bias} partons. Using newly-devised analysis techniques we find that low-$Q^2$ parton fragmentation is accessible down to hadron $p_t$ = 0.35 GeV/c for both hadrons of a correlated pair. Jet morphology for low-$Q^2$ partons requires a more general treatment of fragment $p_t$ distributions and angular correlations. Fragment distributions on transverse rapidity $y_t$ are `infrared safe' and exhibit interesting systematic behaviors which can be compared with the MLLA. Jet angular correlations show a dramatic asymmetry about the thrust axis at low-$Q^2$ (up to 2:1 aspect ratio, with larger width in the azimuth direction), possibly related to nonperturbative details of soft parton collisions. These measurements of p-p correlations provide an important reference for the study of {\em in-medium modification} of parton scattering and fragmentation in heavy ion collisions.

This work was supported in part by the Office of Science of the U.S. DoE under grant DE-FG03-97ER41020.


\end{document}